\documentclass[slac_one]{revtex4}
\usepackage{graphicx}
\usepackage{fancyhdr}
\begin{document}

\title{Recent Progress in Lattice QCD} 

%

\author{Tetsuya Onogi}
\affiliation{YITP, Kyoto University, Kyoto 606-8502, Japan}

%

\begin{abstract}
In recent years, reallistic unquenched QCD simulations have 
been carried out with various lattice actions. 
In this report, I explain the progress in theory and algorithms 
and some of the physics results.

\end{abstract}

\maketitle

\thispagestyle{fancy}


\section{INTRODUCTION} 
The hadronic states in QCD have various energy scales: the smallest scales are 
the light pseudoscalar meson masses $m_{\pi}$ and $m_K$, 
and the medium energy scales 
other light hadronic masses of order $\Lambda_{\rm QCD}$, 
and the larger energy scales are heavy hadron masses $m_D$ and $m_B$.
 These mass scales ranges
from 100 MeV to 5 GeV, giving a hierarchy of order 50. 
If one wants to fully describe such a system with completely 
controlled systematic errors, 
one should ideally carry out lattice simulations 
1) with sufficiently
large volume $V=L^4$ and small pseudoscalar meson masses $m_{\rm PS}$ 
to cover the chiral dynamics of the $\pi$ meson and 
2) with sufficiently 
large heavy quark mass $m_Q$  and large cutoff scale $a^{-1}$ in order
to cover the heavy meson mass scale. 
However, in reality, due to the
limitation of the present computer power  our lattice simulation can 
marginally cover the energy scales as 
\begin{eqnarray}
L^{-1} < m_{\pi} < m_{\rm PS}  < m_K < \Lambda_{\rm QCD} < m_c < m_Q <
a^{-1} < m_b . 
\end{eqnarray}
Even if one cannot totally cover the energy scale, one can 
extrapolate the simulation result to the physical result with 
the help of the chiral perturbation theory and the heavy meson
effective theory. Therefore, the real practical problem is to achieve
lattice computations with larger 
volume, smaller light quark mass, and finer lattice as much as
possible so that the kinematic parameters lie with the range where the 
effective theory descriptions are valid.  

In addition, in order to obtain the hadronic matrix element of the
operator ${\cal O}$ with high precision, 
we need to determine the renormalization factor $Z_{\cal O}$ 
very accurately since the matrix element in the continuum theory 
can be related to that on the lattice as 
\begin{eqnarray}
\langle {\cal O}^{cont}(\mu) \rangle
\equiv 
\lim_{a\rightarrow 0}
Z(a\mu,g_0(a)^2) \langle {\cal O}^{lat}(a) \rangle.
\end{eqnarray}
In recent years, there have been several new developments 
both in theoretical formulations and algorithms. With these 
developments, recent lattice simulations can cover much wider energy 
scale so that the extrapolation of the quark masses to their physical 
values can be carried out reliably with the help of the low energy
effective theories.
\subsection{Lattice fermion actions}
Let us now comment on various lattice fermion actions.
The staggered fermion or improved staggered fermions (AsqTad, HISQ) 
have been extensively used for large scale unquenched simulations. 
The biggest 
advantages are the small simulation cost due to the small degrees of 
freedom and the exact partial chiral symmetry 
at finite lattice spacing. Since it has extra 'taste' degrees of 
freedom, square or quartic rooted trick are needed. The Wilson fermion 
or O(a)-improved Wilson fermion (=Clover fermion) are now widely used 
in recent unquenched simulations. The biggest advantage is that this 
action is theoretically most simple and the computational cost is
reasonable. Although the chiral symmetry is violated at finite lattice 
spacing, it can be recovered in the continuum limit.  
The Domain-Wall fermion ~\cite{DW} is a Wilson fermion in 5 dimensions 
with a mass term which has kinks in the 5-th dimensions. 
The chiral symmetry violations are exponentially suppressed as a 
function of the lattice extent in the 5-th dimension
$N_5$. It is $N_5$ times more costly than the conventional Wilson fermion.
The overlap~\cite{overlap} fermion has an
exact chiral symmetry ~\cite{Luscher:1998pq} on the lattice. 
This action is very costly but still feasible by the new method 
to fix topology during the hybrid Monte Carlo simulation. 
The twisted mass Wilson (tmWilson) is the Wilson fermion with chirally 
rotated mass. It has a partial exact chiral symmetry but partial
vector symmetry is violated, instead. At maximal twist, this action is
automatically O(a)-improved.
\subsection{New algorithms in unquenched simulations}
It is known that the simulation cost grows towards 
smaller quark mass, larger volume, and finer lattice spacing. 
In addition to the obvious effect from simply having more lattice
points, there are effects from more iterations in Dirac matrix
inversion, smaller step size in keeping the acceptance rate constant,
and larger Monte Carlo trajectories to compensate longer autocorrelations.
For example, the empirical formula ~\cite{berlinwall} for $N_f=2$ QCD
with O(a)-improved Wilson 
fermion based on the studies by CP-PACS and JLQCD collaborations with
conventional algorithm 
is
\begin{eqnarray}
\mbox{cost}\left[\mbox{Tflops}\cdot \mbox{years}\right]
&=& C \left[\frac{\#\mbox{conf}}{1000}\right] 
\cdot \left[\frac{0.6}{m_{\pi}/m_{\rho}}\right]^6
\cdot \left[\frac{L}{\mbox{3 fm}}\right]^5
\cdot \left[\frac{0.1\mbox{fm}}{a}\right]^7
\end{eqnarray}
with $C\simeq 2.8$ . From this formula, one can estimate that 
a typical simulation of 1000 configurations with $a=0.1$fm, $L=3$ fm
would cost 25 Tflops years for $m_{\pi}=300$ MeV and more than 600
Tflops years for physical pion mass. For this reason, it has been
thought that the staggered fermion is the only feasible choice for 
unquenched QCD simulations for $m_{\pi} \simeq 300$ MeV with Tflops 
machine. 

One of the most striking developments in lattice QCD is the 
proposal of new algorithms to reduce the simulation cost of unquenched
QCD simulation. The new algorithms are the Hybrid Monte Carlo 
based on the combination of the 'preconditionning' (mass preconditionning
~\cite{massprec1,massprec2} or domain decomposition 
~\cite{luscher,kennedy}) to reduce the cost of the Dirac operator 
inversion and 'multi-time scale' in Molecular dynamics to reduce the 
frequency of the Dirac operator inversion in time-step.
It was found that these new algorithms can give rise to significant
cost reductions of factor 20-30. Therefore, the study of 
O(a)-improved Wilson fermions has also become quite feasible with 
O(1-10) Tflops machines. The same algorithm also enables large scale 
simulations of Domain-wall or overlap fermions, although one must
restrict oneself to work on a single relatively coarse lattice spacing
with the present computer.  
\begin{table}[t]
\begin{center}
\caption{Large scale simulation projects}
\begin{tabular}{|l|l|l|l|l|l|}
\hline
Group & Fermion Action  & $n_f$ & a(fm) & L(fm) & $m_{\pi}$(MeV) \\
\hline
MILC~\cite{milc07}  & Asqtad   & 2+1 & 0.06, 0.09, 0.12,0.15 & 2.4-3.4 & $\geq$ 240\\
PACS-CS~\cite{Aoki:2008sm} & Clover  & 2+1 & 0.07, 0.09, 0.10  & 3       & $\geq$ 150\\
BMW~\cite{BMW}    & Clover      & 2+1 & 0.065,0.085,0.125     & 2-4     & $\geq$ 300\\
CERN~\cite{del06,del07}  & Wilson   & 2   & 0.052,0.072,0.078     & 2.5-3.5 & $\geq$ 300\\
CLS/Alpha~\cite{CLS} & Clover      & 2   & 0.04, 0.06, 0.08,     & 1.9-2.4 & $\geq$ 300\\
ETMC~\cite{urbach07}  & tmWilson    & 2   & 0.07, 0.09, 0.10      & 3       & $\geq$ 300\\
\hline
RBC/UKQCD~\cite{rbcukqcd08} & Domain-wall & 2+1 & 0.086,0.12  & 3       & $\geq$ 330\\
JLQCD~\cite{Noaki:2008iy}  & Overlap     & 2   & 0.12                  & 2       & $\geq$ 300\\
          &             & 2+1 & 0.11                  & 2       & $\geq$ 300\\
\hline
\end{tabular}
\label{tab:projects}
\end{center}
\end{table}
\subsection{Nonperturbative renormalization}
Another important theoretical methods is the nonperturbative 
renormalization.There are several different methods.

The first method is a simple method using the PCAC relation
$\Delta_{\mu} A_{\mu} = 2 m_q P$.
If the axial current is exactly conserved on the lattice 
(ex. staggered, tmWilson, Overlap), one can use PCAC relation to 
obtain the decay constant of the pseudoscalar meson with
nonperturbative accuracy as 
$\displaystyle{f_{\rm PS} = \lim_{a\rightarrow 0} 
\frac{\langle 0 | 2 m_q P  | 0 \rangle}{m_{\rm PS}^2}}$. 
The second method is the nonperturbative renormalization 
by the Regularization Independent Momentum (RI-MOM) scheme proposed
by Martinelli et al.~\cite{Martinelli:1994ty}. This is a scheme
defined by the quark-gluon 
amplitudes in Landau gauge with large off-shell momentum in Euclidean
region. 
Schrodinger functional (SF) scheme proposed by the Alpha
collaboration~\cite{luscher92,luscher93_SU2,Luscher94}. This scheme is
defined the amplitudes in a box  
with boundaries with physical box size $L^4$ under the external 
gauge potential introduced by the boundary values of the gauge field.
\section{LIGHT FLAVOR PHYSICS}
\subsection{Quark mass dependence of the pseudoscalar masses and decay
constants} 
\begin{figure*}[t]
\centering
\includegraphics[width=80mm]{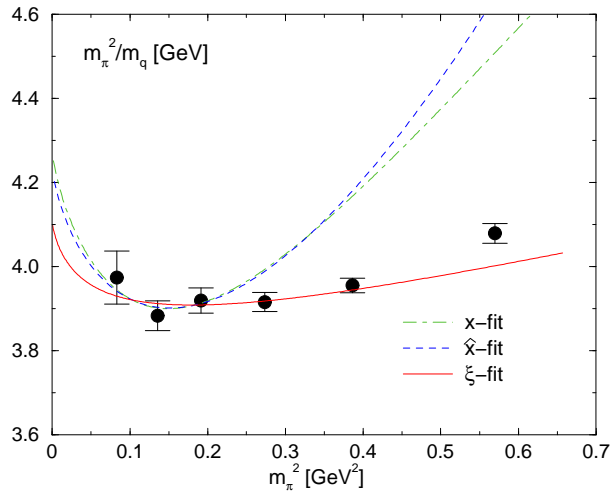}
\caption{ChPT fit with $x$-expansion and $\xi$-expansion. 
$x$-expansion makes the convergence of the ChPT fits better.}
\label{fig:mpi_mud_JLQCD}
\end{figure*}

Next-to-Leading Order (NLO) SU(2) Chiral Perturbation Theory (ChPT)
predicts the quark mass dependences of the pseudoscalar masses and
decay constants as 
\begin{eqnarray}
m_{\rm PS}^2 
= 2 m_q B \left[ 1 + \frac{2 m_q B}{16\pi^2 f^2}\left(\bar{l}_3 +
\ln(\frac{2m_q B}{m_{\pi}})\right)\right],
& f_{\rm PS}
= f \left[ 1 + \frac{2 m_q B}{8\pi^2 f^2}\left(\bar{l}_4 -
\ln(\frac{2m_q B}{m_{\pi}})\right)\right],
\end{eqnarray}
where $m_{\pi}$= 139.6 MeV. 
\begin{figure*}[t]
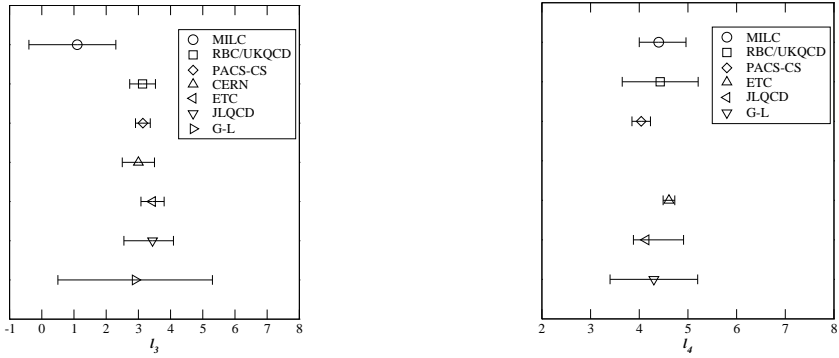

\vspace{1cm}
\hspace{-3cm}\includegraphics[width=40mm]{l3.eps}
\hspace{3cm}\includegraphics[width=40mm]{l4.eps}
\caption{Low energy constants $\bar{l}_3,\bar{l}_4$ from NLO SU(2)
ChPT fits of the lattice data from various groups. For comparison, 
phenomenological estimates by Gasser and Leutwyler are also plotted. } 
\label{fig:NLO_ChPT}
\end{figure*}
The ratio of the decay constant is also obtained by various groups.
Table~\ref{tab:fK_fpi} lists the ratio $f_K/f_{\pi}$ in 2 and 2+1
flavor QCD from various groups.
\begin{table}[t]
\begin{center}
\caption{SU(3) breaking in the pseudoscalar decay constant.}
\begin{tabular}{lllll}
\hline
Group     &$n_f$& Action      & input               &$f_K/f_{\pi}$ \\
\hline
MILC      & 2+1 & Asqtad      &$m_{\Upsilon,\pi,_K}$ &1.202(3)($^{+6}_{-13}$)\\
HPQCD/MILC& 2+1 & Asqtad+HISQ &$m_{\Upsilon,\pi,K}$  &1.189(7)\\
PACS-CS   & 2+1 & Clover      &$m_{\Xi,\pi,K}$       &1.19(2)\\
BMW       & 2+1 & Clover      &$m_{\Omega,\pi,K}$    &1.19(1)(1)\\
RBC/UKQCD & 2+1 & Domain-wall &$m_{\Omega,\pi,K}$    &1.22(2)(6)\\
JLQCD     & 2+1 & Overlap     &$f_{\pi},m_{\pi,K}$   &1.20(3)\\
ETMC      & 2   & tmWilson    &$f_{\pi},m_{\pi,K}$   &1.196(13)(7)(8)\\
\hline
\end{tabular}
\label{tab:fK_fpi}
\end{center}
\end{table}
A remarkable observation was made by the 2+1 flavor simulation 
PACS-CS collaboration~\cite{Aoki:2008sm}, where the lightest quark
mass is as small as  
the physical up and down quark masses. They find that the non-analytic 
behavior as functions of $m_{ud}$ arising from the chiral log can be 
clearly reproduced for $m_{\pi}^2/m_{ud}$ and $f_K/f_{\pi}$. 
RBC/UKQCD collaboration~\cite{rbcukqcd08} found that the NLO partially
quenched chiral 
perturbation theory (PQChPT) with the expansion parameter
$x=m_{\pi}^2/(4\pi f)^2$ ($x$-expansion)fits the data only for
$m_{\pi} <450$ MeV.
While JLQCD collaboration finds similar results for ChPT, they also find 
that the NLO ChPT with expansion parameter $\xi=m_{\pi}^2/(4\pi
f_{\pi})^2$ ($\xi$-expansion) makes the convergence better for heavier
quark masses  
and NNLO ChPT with $\xi$-expansion can nicely describe the lattice data 
up to the strange quark mass regime as shown in
Fig.~\ref{fig:mpi_mud_JLQCD}~\cite{Noaki:2008iy}.
\subsection{$B_K$}
Indirect CP violation in the K meson system $\epsilon_K$ is one of 
the most crucial quantities to test the standard model and the physics
beyond. 
The experimental value is determined with high accuracy as
\begin{eqnarray}
|\epsilon_K|= (2.233\pm 0.015)\times 10^{-3}.
\end{eqnarray}
Theoretically this quantity is described as
$|\epsilon_K|=  f(\rho,\eta) \times C(\mu) \times B_K(\mu)$.
Here, $f(\rho,\eta)$ is a factor which depends on the CKM matrix
elements, $C(\mu)$ is the Wilson coefficient from short-distance
QCD corrections and $B_K(\mu)$ is the bag parameter defined as
\begin{eqnarray}
B_K(\mu) = \frac{\langle K^0 | \left[\bar{d}\gamma^{\mu}(1-\gamma_5) s
\bar{d}\gamma_{\mu}(1-\gamma_5) s \right](\mu) | \bar{K}^0
\rangle}{\frac{8}{3} f_K^2 m_K^2}. 
\end{eqnarray}
The main problem in unquenched lattice calculations is the 
mixing of operators with wrong chiralities or tastes. For Wilson-type
fermions  
there exists mixing with operators of the same dimension having wrong
chirality, Since they cannot be extrapolated away in the continuum
limit, the only way to remove the contamination is to determine the 
counter term nonperturbatively. However, due to the chiral enhancement 
of the operators with wrong chirality, it is quite difficult to
control the systematic error. The operator mixing for staggered
fermion is also quite complicated and only perturbative
renormalization exists. The overlap fermion is free from operator
mixing owing to the exact chiral symmetry, while the operator 
mixing for the domain-wall fermion is exponentially suppressed 
and practically under control. In fact,  for both cases studies 
with nonperturbative renormalization in RI-MOM scheme, one cannot 
see visible effect of the the operator mixing. 
For tmWilson, due to the parity odd nature and the cs symmetry 
the operator mixing can be prohibited for 2-flavor QCD. Therefore,
tmWilson is another promising approach to the precise determination 
of $B_K$.  However, for 2+1 flavors it is difficult to full realize
the O(a)-improvent, no mixing and non-degenerate $(m_s,m_d)$ quark 
simultaneously. 

HPQCD collaboration studied $B_K$ on 2+1 flavor MILC configuration 
with Aqstad (staggered ) sea quark with lattice spacing $a=0.125$ fm
~\cite{Gamiz:2006sq}. They employ two different improved staggered
actions (Asqtad and HYP) 
for the valence quark. The calculation was carried out by degenerate
valence quarks for the K meson $(m_1,m_2)=m_s/2$ while the sea quark
is chosen to be $0.2 m_{sea}, 0.4 m_s$. Chiral extrapolation 
is performed by linear fit. The renormalization of $\Delta S=2$
operator is obtained by perturbation at 1-loop. They find 
$\hat{B} = 0.83(18)$, where the dominant error comes from the 
unknown higher order perturbative correction of the renormalization 
factor of $O(\alpha^2)\simeq 20\%$.
RBC/UKQCD collaborations computed $B_K$ with domain-wall fermion 
in 2+1 flavor QCD at lattice spacing $a=0.11$ fm~\cite{rbcukqcd08}. They have 3 points 
for the sea quark and 7 combinations of valence quark masses 
$(m_1,m_2)$. They fit the data with next-to-leading order partially
quenched chiral perturbation theory (NLO PQChPT). The renormalization 
factor is determined nonperturbatively by RI-MOM scheme. They obtain 
$\hat{B} = 0.720(13)_{\rm stat.}(17)_{\rm sys.}$, where the dominant 
error comes from the discretization error of $O(a^2)\simeq 4\%$.
JLQCD collaborations study $B_K$ with overlap fermion 
in 2 flavor QCD at lattice spacing $a=0.12$ fm on physical volume with
$L=2fm$~\cite{Aoki:2008ss}. They have 4 points for the sea quark and 10 combinations of
valence quark masses  
$(m_1,m_2)$. They fit the data with NLO PQChPT. The renormalization 
factor is determined nonperturbatively by RI-MOM scheme. They obtain 
$\hat{B} = 0.734(5)_{\rm stat.}(50)_{\rm sys.}$, where the dominant 
error comes from the finite size effect of order  $5\%$.

It should be noted that the long standing operator mixing problem 
is solved with the advent of Ginsparg-Wilson fermions (domain-wall, 
overlap) with (almost) exact chiral symmetry. Thus the above studies 
are the real start of the precision study of $B_K$ for which 
significant progress will be expected near future. 
RBC/UKQCD collaboration are planning to work at finer lattice spacing 
$a=0.09$ fm to reduce the discretization error. JLQCD collaborations 
are studying 2+1 flavor QCD for physical volumes of 2 fm and 3 fm.
ETMC and other groups are studying on mixed action approach 
where they employ the Ginsparg-Wilson valence quarks on gauge 
configuration with tmWilson or staggered dynamical quarks. 
This can also be an economical and promising approach, with the 
help of mixed action PQChPT so that one can even take the continuum 
limit with the existing 2+1 flavor gauge configurations.

More precise determination of $B_K$ are important in view of the 
recent tension in $\sin(2\phi_1)$ versus the determination 
of unitarity triangle by $\epsilon_K$ and $\Delta
m_{B_s}/\Delta_{B_d}$ as pointed out by Lunghi and Soni~\cite{Lunghi:2008aa}
or Buras and Guadagnoli~\cite{Buras:2008nn}.

\subsection{Nucleon sigma term}
Nucleon sigma term is the finite quark mass effect of the nucleon mass
and is defined by the following matrix element
\begin{eqnarray}
\sigma_{\pi N} &=& m_{ud} \langle N | \bar{u}u + \bar{d}d | N \rangle,
\end{eqnarray}
which contains both the valence and the sea quark contributions.
The sea quark contribution also arise from the strange quark and 
it is often parameterized by the ratio $y$ defined as 
\begin{eqnarray}
y &=&  \frac{2 \langle N | \bar{s}s | N\rangle }
            {\langle N | \bar{u}u + \bar{d}d | N\rangle}.
\end{eqnarray} 
The sigma term and the strange quark content can be determined 
using the quark mass dependence of the nucleon mass.
The Feynman-Hellman theorem tells that the quark 
mass dependence can be related to the sigma term as 
\begin{eqnarray}
\frac{\partial m_N}{\partial m_{\rm val}}
= \langle N | \bar{u}u + \bar{d}d | N \rangle_{\rm conn}, 
&\frac{\partial m_N}{\partial m_{\rm sea}}
= \langle N | \bar{u}u + \bar{d}d | N \rangle_{\rm disc}, 
\end{eqnarray}
where ``conn'' and ``disc'' denotes contributions to the sigma terms. 
The disconnected contribution is nothing but $2 \langle N | \bar{s}s | N \rangle$
when the sea quark mass is equal to strange quark mass. 

JLQCD collaboration studied the quark mass dependence of the nucleon mass 
in 2-flavor QCD simulation with dynamical overlap fermion~\cite{Ohki:2008ff}. 
Fitting the nucleon mass for the combination of 6 sea quark masses and
9 valence quark masses by partially quenched baryon chiral
perturbation theory,
\begin{figure*}[t]
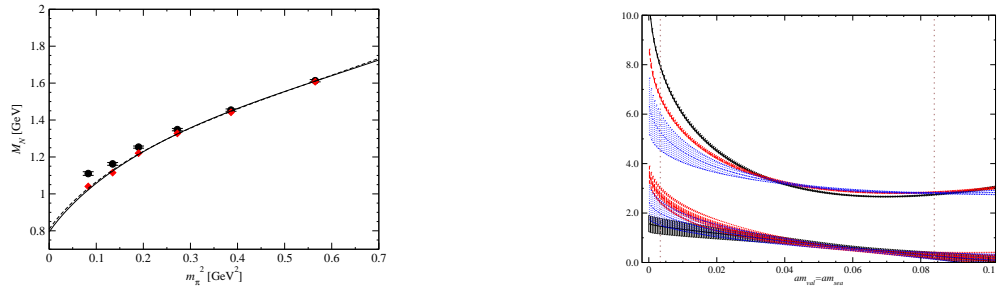

\vspace{5mm}
\hspace{-3cm}\includegraphics[width=50mm]{FSE_FIT0.eps}
\hspace{3cm}\includegraphics[width=50mm]{val_vs_sea.eps}
\caption{Left panel shows the quark mass dependence of the nucleon mass in 2-flavor QCD.
Right panel shows the connected and disconnected contributions to the
sigma term.}\label{fig:sigma}
\end{figure*}
They found that the by fitting the data with BChPT in finite volume 
finite size effect for the baryon mass in $(2fm)^4$ box is
significant. After correcting the finite size effect, they obtain
\begin{eqnarray}
\sigma_{\pi N}&=&52(2)(^{+20}_{-2})(^{+5}_{-0}) \mbox{ MeV}.
\end{eqnarray}

Fitting the nucleon mass data with PQChPT and differentiating 
the nucleon mass with valence and sea quark masses. They found that 
the disconnected contribution is much smaller than connected one.
As a result the semi-quenched estimate of the y-parameter is
\begin{eqnarray}
y&=& 0.030(16)(^{+6}_{-8})(^{+1}_{-2}).
\end{eqnarray}

Previous lattice results  y-parameters using Wilson-type fermions 
are 0.66(15), 0.36(3) for quenched QCD and 0.59(13) for 2-flavor QCD, 
which are significantly larger than the above result. 
It was pointed out that the the additive quark mass
shift with Wilson-type fermion can give a significant contamination 
to the disconnected contribution. After subtracting this artifact, 
UKQCD obtained $y=-0.28(33)$ in 2-flavor QCD. Since they subtract 
the power-divergent contribution numerically, the result suffer from 
large statistical error. In contrast, the study with overlap fermion 
by JLQCD do not suffer such artifact owing to the exact chiral symmetry. 

The exact chiral symmetry on the lattice enables novel studies of 
chiral dynamics. It can also open up new directions which has not 
been possible in other approaches. Examples of such studies are given
by JLQCD collaboration~\cite{Shintani:2008ga, Shintani:2008qe, 
Aoki:2008tq, Fukaya:2007pn, Aoki:2007pw, Aoki:2007ka,
Fukaya:2007yv, Fukaya:2007fb}.      
\section{CHARM QUARK PHYSICS}
\subsection{D meson decay constants}
Since the charm quark mass is of order 1 GeV, 
the discretization errors are larger than those for up, down, and
strange quark if one take the same approach used for the light quark. 
In order to make a reliable extrapolation should make the lattice 
one has to employ either finer lattices or improved action.

The first choice is made by the Alpha collaboration. They carry
out a quenched study the $D_s$ meson decay constant on lattices  
with lattice spacings $a=0.04,0.06,0.08, 0.1$ fm, which correspond to
the discretization errors of $(am_c)^2 \simeq 0.04,0.09,0.16,0.25$
~\cite{Juttner:2003ns}. Using the O(a)-improved Wilson fermion and
nonperturbative renormalization in SF-scheme, they obtain
\begin{eqnarray}
f_{D_s}^{n_f=0}=252(2) \mbox{ MeV}.
\end{eqnarray}
They same strategy is in progress for $n_f=2$ QCD by the Coordinated
Lattice Simulation(CLS)/Alpha collaboration. This is a quite promising 
approach but requires dedicated effort.
 
Improved fermion actions are also effective in reducing the
discretization error. One of the popular approaches is the fermilab
formalism and other related formalism in which one make a
mass-dependent re-interpretation of $O(a)$-improved Wilson
fermion. While it is possible to achieve non-perturbative improvement,
mass-dependent errors of the leading operators are removed at tree and
1-loop in practice. The dominant short distance error $\Delta$ for the
charm quark system is
\begin{eqnarray}
\Delta
 = c_0 \alpha_s^2 
  +c_1 \alpha_s (a\Lambda_{\rm QCD}) (a m_c)
  +c_2 (a\Lambda_{\rm QCD})^2 + \cdots
\end{eqnarray}
Highly Improved Staggered Quark (HISQ) is another improved action
~\cite{Follana:2006rc}. Since it is based on the standard Symanzik
improvement which treats  
the discretization errors in powers of $(ap)^2$ and $(am_q)^2$, 
standard nonperturbative renormalization technique also applies.
The advantage of this approach is that the discretization errors 
are restricted due to the exact chiral symmetry of the staggered 
fermion. In general, after carrying out nonperturbative
renormalization the dominant short distance error $\Delta$ for the 
charm quark system can be expanded in powers of $(am_c)^2$.
In HISQ action, they remove the leading $O(a^2)$ error and the taste changing 
interaction at tree level so that the dominant short distance then 
becomes
\begin{eqnarray}
\Delta
 =  c_1 (am_c)^4 
  + c_2 \alpha_s (am_c)^2+ \cdots .
\end{eqnarray}
In principle, HISQ has nonperturbative accuracy in the continuum
limit. Whether typical lattice spacings $a\simeq 0.10$ fm are 
small enough to have a good scaling for charm quark system as expected
from naive order counting should be explicitly examined.

\begin{figure*}[here]
\centering
\includegraphics[width=70mm]{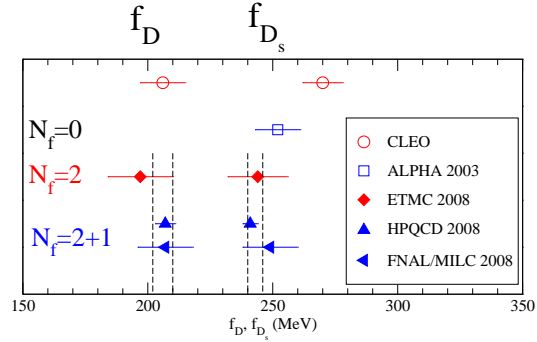}
\caption{$D$ meson decay constants}
\label{fig:fD}
\end{figure*}

HPQCD collaboration studied charm quark system on MILC lattices 
with 2+1 dynamical Asqtad quark using HISQ action for the valence
quark with the lattice spacings $a=0.09,0.12,0.15$
fm~\cite{Follana:2007uv}. The chiral and continuum extrapolation for
the pseudoscalar and  
masses and decay constants is carried out by the global fit of 
the data using the ChPT and heavy meson effective theory formula 
including $O(\alpha_s a^2)$ and $O(a^4)$ errors. 
The scale is set by the upsilon spectrum and $m_{ud}$, $m_s$ and 
$m_c$ are determined from $m_{\pi}$, $m_K$ and $m_{\eta_c}$.
Since there are no more free parameter, 
while $f_{D_s}$ and $f_{D_d}$ are pure predictions of theory, 
it should be noted that $f_{\pi}$, $f_K$, $m_{D_s}$, $m_{D_d}$ and
$m_{D_s}$ can also give non-trivial consistency check.
It turns out mixed action approach with HISQ valence and Asqstad sea 
quark can control the discretization error in the charm quark system 
nicely. They showed that the $D$ and $D_s$ meson masses are reproduced 
in the continuum limit. 
The pseudoscalar decay constant are computed with nonperturbative
renormalization by AWTI, the decay constants $f_{\pi}$ and $f_K$ 
agree with experiment very well. Using the same technique 
they obtain $f_{D_d}$ and $f_{D_s}$ with 1-2\% error which is dominated 
by the continuum and chiral extrapolation. 
\begin{eqnarray}
f_{D_d}=207(4) \mbox{ MeV}, & f_{D_s}=241(3) \mbox{ MeV}, 
& f_{D_s}/f_{D_d}=1.164(11). 
\end{eqnarray}
%
%
Fermilab/MILC collaboration also studied the D meson decay constants 
on the same 2+1 flavor MILC
configurations~\cite{Bernard:2008dn, Bernard:2009wr}. They used AsqTad
action  
for the  light valence quark and fermilab action for the charm quark.
The data are fitted by the PQstaggered ChPT which include full NLO and 
analytic part of the NNLO contributions and $O(a^2)$ errors. 
The heavy-light axial vector current is renormalized by partially
nonperturbative renormalization in which they use the renormalization 
factors of the heavy-heavy and light-light vector currents 
$Z_{V_4}^{QQ}$,  $Z_{V_4}^{qq}$ and evaluate the remaining perturbative correction  $\rho_{A_4}^{Qq}$ 
\begin{eqnarray}
Z_{A_4}^{Qq}&=& \rho_{A_4}^{Qq} \sqrt{Z_{V_4}^{QQ} Z_{V_4}^{qq}}
\end{eqnarray}
They expect that vector current renormalization factors capture 
the dominant contribution from the wavefunction renormalization.
Their preliminary result is
\begin{eqnarray}
f_{D_d}=207(11) \mbox{ MeV}, & f_{D_s}=249(11) \mbox{ MeV}, 
& f_{D_s}/f_{D_d}=1.20(3). 
\end{eqnarray}
ETMC studied the D meson decay constants for $n_f=2$ QCD using 
dynamical tmWilson with lattice spacing $a=0.67,
0.86, 0.10$ fm~\cite{Blossier:2008dj}. Since they work at maximal
twist which guarantees  
automatic $O(a)-$ improvement, the leading discretization error is
$O(a^2)$. They take degenerate ud quark masses for the sea and valence
quark in the range of $m_{ud}=(0.2-0.4)m_s$. The data are fitted with 
SU(2) heavy meson ChPT with $O(a^2)$ term and nonperturbative
renormalization by AWTI is used for the heavy-light axial vector
current. They obtain the preliminary result as 
\begin{eqnarray}
f_{D_d}=197(16) \mbox{ MeV}, & f_{D_s}=244(12) \mbox{ MeV}, 
& f_{D_s}/f_{D_d}=1.24(5). 
\end{eqnarray}
Fig.~\ref{fig:fD} shows the comparison of the $D$ meson decay
constants. It can be seen that $f_{D_d}$ from HPQCD is in agreement 
error with experiment, while $f_{D_s}$ has 3 $\sigma$ deviation from
CLEO result. Two recent preliminary results from Fermilab/MILC and
ETMC give consistent result. It was pointed out by Becirevic et al. 
~\cite{Becirevic}
that possible experiment error by misidentifying $D\rightarrow
l\nu\gamma$ decay could only give an effect smaller than a few percent.  
As was pointed out by the lattice talk by Fermilab group, 
the 3 $\sigma$ deviation is dictated by the error from CLEO 
even if we enlarge the error from HPQCD by factor 3, the deviation 
still 2.8 $\sigma$. Therefore the $f_D$ puzzle still remains. Further 
theoretical and experimental studies are required.
\vspace{1cm}
Combining HPQCD($n_f=2+1$), FNAL/MILC($n_f=2+1$), and ETMC($n_f=2$)
results where I include additional 5\% error for $n_f=2$ results 
my world average is 
\begin{eqnarray}
f_{D_d}=206(4) \mbox{ MeV}, & f_{D_s}=243(3) \mbox{ MeV}, 
\end{eqnarray}

\subsection{Charm quark mass}
Another important progress is the precise determination of the charm.
quark mass. HPQCD and Chetyrkin, Kuhn, Steinhauser and Sturm 
~\cite{Allison:2008xk, Davies}
determined $m_c$ and $\alpha_s$ from the moments the pseudoscalar 
correlators 
\begin{eqnarray}
G_n \equiv \sum_t \left(\frac{t}{a}\right)^n G(t),&
G(t) \equiv a^6 \sum_{\vec{x}}(a m_0^c)^2 
\langle 0 | j_5(\vec{x},t) j_5(\vec{0},0) | 0 \rangle, &
j_5\equiv \bar{\psi}_c \gamma_5 \psi.
\end{eqnarray}
They match the continuum limit of the lattice calculation of $G_n$ 
with the continuum perturbation theory of $G_n$ at 4-loop.
The lattice calculation was performed with 
HISQ valence charm based MILC lattice for $n_f=2+1$ QCD 
at lattice spacings $a=0.09, 0.12, 0.15$ fm.

In order to reduce the discretization error, modified ratios 
$R_n$ are defined
\begin{eqnarray}
R_4 \equiv G_4/G_4^{(0)}, & 
R_n \equiv \frac{a m_{\eta_c}}{2 a m^0_c}
\left( G_n/G_n^{(0)} \right)^{1/(n-4)}  \mbox{ for } n\geq 6,
\end{eqnarray}
where $G_n^{(0)}$ are the n-th moments on the lattice at the tree
level. 
The lattice spacing dependence of the moments 
can be fitted nicely with quadratic functions in $a^2$. 
Averaging for $n=6,8,10$ one can determine the charm quark mass 
and the strong coupling as
\begin{eqnarray}
m_c^{\overline{MS}}(3GeV) = 0.986(10), & & 
\alpha^{\overline{MS}}(M_z,n_f=5) = 0.1174(12).
\end{eqnarray}
The key ingredients are the $O(a^2)$ improvement and the 
exact chiral symmetry, which can control the discretization 
error and also allow nonperturbative renormalization by 
AWTI. 
%
%
\subsection{B PHYSICS}
Even with the finest lattice spacing for the present unquenched
lattice calculation $a=0.04$ fm, the bottom quark mass in lattice 
unit is in the range of $a m_b \simeq 1$. Therefore, HISQ action 
completely looses control over the discretization error. 
Therefore, the only practically available methods to study B physics 
are either to use the effective action such as NRQCD or fermilab a
action or to make a very precise computation in both the static limit 
and in the charm quark mass regime and make an interpolation.
While the former approach are taken by HPQCD and FNAL/MILC
collaborations and has already produced many results in the past few 
years, there has also been remarkable progress in the latter approach, 
where serious feasibility tests of  the theoretical method have been 
carried out in quenched QCD. 

Alpha collaboration worked on the lattice HQET with nonperturbative 
accuracy~\cite{DellaMorte:2007ij, Della Morte:2006cb}. 
Their idea is to make a nonperturbative matching of the HQET 
and QCD at very fine lattices in small volume and then evolve the
lattice HQET to coarse and larger lattice by step scaling. They could 
also work on $1/m_b$ correction in HQET. 

Rome II group used step scaling using $O(a)$-improved Wilson fermion 
with nonperturbative
renormalization~\cite{deDivitiis:2003wy,deDivitiis:2003iy}. 
They make nonperturbative
calculations of B meson directly at bottom quark mass in a small 
volume $L=0.4$ fm at very fine lattice spacing. Then evaluate 
the finite volume correction with coarser lattice by extrapolations
from smaller quark mass regime. 

Guazzini, Sommer and Tantalo combined the above two methods and 
used the static results to evaluate the finite volume correction 
by interpolation~\cite{Guazzini:2007ja}. 

The quenched QCD results are summarized in Table~\ref{tab:Rome_Alpha}.
\begin{table}[t]
\begin{center}
\caption{Quenched QCD results of $f_B$ and $m_b$}
\begin{tabular}{|l|l|l|l|}
\hline 
Group & Method & $f_{B_s}$ (MeV) & $m_b^{\overline{MS}}(m_b)$ (GeV)\\
\hline 
Alpha ~\cite{DellaMorte:2007ij, Della Morte:2006cb}
 & Static + $1/m_b$ & 193(7)& 4.35(5)
\\
Rome II ~\cite{deDivitiis:2003wy,deDivitiis:2003iy}
& finite volume step scaling  & 192(6)(4) & 4.33(10)\\
Guazzini et al. ~\cite{Guazzini:2007ja}
& combination & 191(6) & 4.42(6)\\
\hline
\end{tabular}
\label{tab:Rome_Alpha}
\end{center}
\end{table}
It is remarkable that one can control the systematic error from the
bottom quark and achieve 2-3\% accuracy in the matrix element. 
$n_f=2$ unquenched QCD studies are in progress by Alpha collaboration.

\begin{figure*}[here]
\centering
\includegraphics[width=70mm]{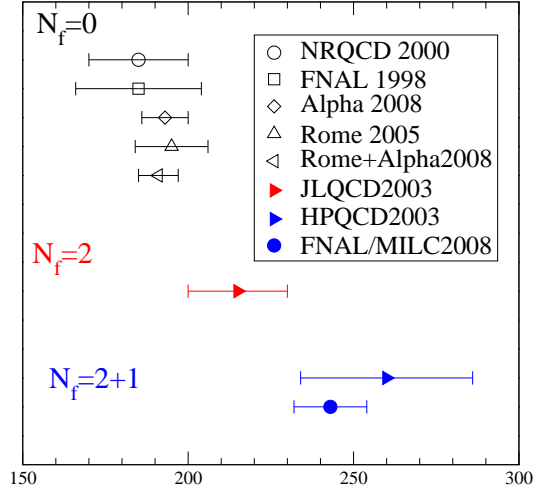}
\caption{$B_s$ meson decay constant.} 
\label{fig:fBs}
\end{figure*}

FNAL/MILC collaboration recently update their $n_f=2+1$ calculation 
of $f_{B_s}$ and $f_{B_d}$~\cite{Bernard:2009wr}. 
Fig.~\ref{fig:fBs} shows the $B_s$ meson 
decay constants in quenched QCD and $n_f=2,2+1$ unquenched QCD 
simulations. Using the input such as $m_{\rho}$ or $r_0=0.5$ fm for 
$n_f=0,2$, one finds that $f_{B_s}$ increase from quenched QCD to 
$n_f=2+1$ in contrast to $f_{D_s}$. 
The SU(3) breaking of the $f_B$ in 2+1 flavor QCD is obtained 
by HPQCD and FNAL/MILC collaboration also reported their result 
at Lattice 2008. Since FNAL/MILC results of $f_B$ are still
preliminary, I just quote their numbers and do not take the 
world average.
\begin{eqnarray}
f_{B_s}/f_{B_d} = 1.20(3) \mbox{ HPQCD}, & &
f_{B_s}/f_{B_d} = 1.25(4) \mbox{ HPQCD}.
\end{eqnarray}
\subsection{$B\rightarrow D^{\ast} l\nu$}
FNAL/MILC collaboration recently made an update of 
$B\rightarrow D^{\ast} l\nu$ form factor determination 
based on 2+1 MILC lattices at lattice spacings $a=0.09, 0.12, 0.15$
fm~\cite{Laiho}. They employ the Asqtad action for the light sea and
valence quarks and the fermilab action for the heavy quark.
They developed a new double ratio method to determined the 
$B\rightarrow D^{\ast} l\nu$ form factor as 
\begin{eqnarray}
|{\cal F(1)}|^2
=\frac{
\langle D^{\ast} | \bar{c}\gamma_j \gamma_5 b | \bar{B} \rangle
\langle \bar{B} | \bar{b}\gamma_j \gamma_5 c | D^{\ast} \rangle}
{\langle D^{\ast} | \bar{c}\gamma_4 c | D^{\ast} \rangle
\langle \bar{B} | \bar{b}\gamma_4 b | \bar{B} \rangle}.
\end{eqnarray}
They carry out the  chiral extrapolation and continuum extrapolation 
using the NLO ChPT and analytic NNLO staggered ChPT formula. 
They obtain 
\begin{eqnarray}
{\cal F}^{n_f=2+1}(1) = 0.921(11)(19)
\end{eqnarray}
as a preliminary result.

Rome II group studied $B\rightarrow D^{\ast} l\nu$ form factor 
in quenched QCD using finite size step scaling method and combined 
with the HQET calculation~\cite{deDivitiis:2008df},~\cite{Tantalo}. 
They used twisted boundary condition to 
get small continuous momentum recoil. As a result they could obtain 
the slope of the $w$ dependence.
They obtain 
\begin{eqnarray}
{\cal F}^{n_f=0}(1) = 0.917(8)(5).  
\end{eqnarray}
This proves that the finite size step scaling method is very
promising also for the form factor calculation. Applications 
to unquenched QCD simulation are awaited.

\subsection{$B\rightarrow \pi l\nu$}
FNAL/MILC  collaboration made a new analysis of 
$B\rightarrow \pi l\nu$ form factor using MILC configuration of 2+1
flavor QCD at $a=0.09, 0.12, 0.15$ fm~\cite{VandeWater}
. 
Parameterizing the the form factors as 
\begin{eqnarray}
f_{\perp}(E_{\pi}) p_i \equiv \langle \pi | V_i | B\rangle
/\sqrt{2m_B}, 
& &
f_{\parallel}(E_{\pi}) \equiv \langle \pi | V_0 | B\rangle /\sqrt{2m_B}
\end{eqnarray}
and using NLO staggered ChPT 
\begin{eqnarray}
f_{\parallel,\perp} &=& c_0
\left[ 1 + \mbox{chiral logs} + m_l 
+ c_2(2m_u+m_s) +c_3 E_{\pi} + c_4 E_{\pi}^2 
+ c_5 a^2 \right] + O(m_q^2, E_{\pi}^3), 
\end{eqnarray}
they carried out the continuum and chiral extrapolation.
They then used the combined $z$ fit of Babar and lattice data 
based on the dispersion relation formula and extracted 
$|V_{ub}|$ as
\begin{eqnarray}
|V_{ub}| = 2.94(35) \times 10^{-3}.
\end{eqnarray}

\subsection{New applications}

The nonperturbative renormalization developed for QCD simulation 
for precise test of the standard model is also useful for exploring 
the beyond standard model physics. Recently, the running coupling 
constant in Schrodinger functional scheme is applied to QCD with 
many flavors. Appelquist \textit{et al.}~\cite{Appelquist:2007hu} 
found an evidence of an infrared fixed 
point in strong coupling regime in QCD with twelve flavors. 
This may serve to construct an example for the walking technicolor.
Application to more realistic models are now in progress.

\section{CONCLUSIONS}

In conclusion, the recent developments of unquenched lattice 
simulation allows us to study the realistic 2+1 flavor QCD. 
A lot of nonperturbative computation in light flavor physics is now 
well under control. Highly improved fermion actions are beginning 
to control the charm physics. The b quark still suffers from
large systematic errors but several promising methods have been 
tested and giving precise determinations of weak matrix elements 
of the B mesons in quenched QCD.
It was found that the exact chiral symmetry is quite useful in some
cases.

Coordinated work to combine advance techniques will lead to few
percent accuracy within few years, including cross-checks with 
different lattice actions.

\begin{acknowledgments}
I would like to thank D.~Becirevic, C.~T.~H.~Davies, A.~S.~Kronfeld,
J~ Laiho, R.~Sommer, N.~Tantalo and A.~Soni for useful discussions.
I would also like to thank P.~B.~Mackenzie, C.~Tarantino,
R.~Van~de~Water for sending their data to prepare my talk. 
Special thanks my colleagues in JLQCD collaboration.
Work is supported in part by the Grantin-
Aid of the Ministry of Education (Nos. 19540286,
20039005).
\end{acknowledgments}

\end{document}